\documentclass[fleqn,usenatbib]{mnras}

\usepackage{newtxtext,newtxmath}

\usepackage[T1]{fontenc}

\newcommand{\fermi}{{\it Fermi}}

\newcommand{\swift}{{\it Swift}}

\newcommand{\ergflux}{\mbox{${\rm \, erg \,\, cm^{-2} \, s^{-1}}$}}
\newcommand{\gm}{$\gamma$}

\DeclareRobustCommand{\VAN}[3]{#2}
\let\VANthebibliography\thebibliography
\def\thebibliography{\DeclareRobustCommand{\VAN}[3]{##3}\VANthebibliography}

\usepackage{graphicx}	
\usepackage{amsmath}	

\title[TXS 1433+205]{TXS 1433+205: The most distant gamma-ray emitting FR~II radio galaxy}

\author[Paliya et al.]{
Vaidehi S. Paliya,$^{1}$\thanks{E-mail: vaidehi.s.paliya@gmail.com (VSP)}
D. J. Saikia$^{1}$
and C. S. Stalin,$^{2}$
\\
$^{1}$Inter-University Centre for Astronomy and Astrophysics (IUCAA), SPPU Campus, Pune 411007, India\\
$^{2}$Indian Institute of Astrophysics, Block II, Koramangala, Bengaluru 560034, Karnataka, India
}

\date{Accepted XXX. Received YYY; in original form ZZZ}

\pubyear{2022}

\begin{document}
\label{firstpage}
\pagerange{\pageref{firstpage}--\pageref{lastpage}}
\maketitle

\begin{abstract}
The orientation of the jet axis to the line of sight of the observer plays a major role in explaining the phenomena observed from blazars and radio galaxies.  In the \gm-ray band, only a handful of radio galaxies have been identified, all being located in the nearby Universe ($z<0.5$). Here we report the identification of 4FGL~J1435.5+2021, associated with TXS~1433+205, as a Fanaroff-Riley type II (FR~II) radio galaxy at a considerably higher redshift of $z=0.748$, thereby making it the most distant \gm-ray detected radio galaxy known as of now. The Very Large Array Sky Survey data at 3 GHz resolves the source morphology into a bright core, a jet and two hotspots, with a total end-to-end projected length between lobe extremities of $\sim$170 kpc.  The optical and radio properties of this enigmatic object suggest it to be a high-excitation FR~II radio galaxy. The multi-wavelength behaviour of TXS~1433+205 is found to be similar to other \gm-ray detected FR~II sources but is at the high luminosity end.  We suggest that the ongoing and upcoming high-resolution radio surveys will lead to the identification of many more high-redshift radio galaxies in the \gm-ray sky, thus allowing comprehensive studies of misaligned relativistic jets.
\end{abstract}




\section{Introduction}
The extragalactic \gm-ray sky as observed by \fermi-Large Area Telescope (LAT) is dominated by active galactic nuclei (AGN) hosting beamed relativistic jets, i.e., blazars. Though in small numbers, \fermi-LAT has also detected significant \gm-ray emission from AGN with misaligned jets, mainly radio galaxies \citep[cf.][]{2020ApJ...892..105A}.  These are relatively \gm-ray faint sources likely due to less extreme Doppler boosting. Nevertheless, they are being extensively studied and surveys are being intensively searched to identify new \gm-ray emitters to study the radiative processes powering relativistic jets and, in general, AGN unification \citep[see, e.g.,][]{2011ApJ...740...29K,2012ApJ...751L...3G,2021ApJ...918L..39P}.  Based on the radio morphology and power, radio galaxies have been further divided as faint edge-darkened Fanaroff-Riley type I (FR~I) and luminous edge-brightened FR~II sources usually showing bright hotspots \citep[][]{1974MNRAS.167P..31F}. More FR~Is have been detected with the \fermi-LAT compared to the FR~II ones \citep[cf.][]{2020ApJ...892..105A}. This could be possibly due to the relatively flat \gm-ray spectra of the former and/or their closer proximity compared to FR~IIs \citep[cf.][]{2022ApJ...931..138F} but it is yet to be fully understood.  

Gamma-ray emitting radio galaxies have mostly been found in the low-redshift Universe \citep[$z<0.5$,][]{2010ApJ...720..912A,2020JHEAp..27...77C}. For example, in the second data release of the fourth catalog of the \fermi-LAT detected AGN \citep[4LAC-DR2][]{2020ApJ...892..105A}, the farthest known radio galaxy is 3C~17 at $z=0.22$. Recently, \citet[][]{2022MNRAS.513..886B} have identified an FR II radio galaxy, IGR~J18249$-$3243, at $z=0.36$ among the unclassified \gm-ray sources listed in the 4LAC catalog.  Identifying \gm-ray emitting radio galaxies at high redshifts is crucial not only to find the most \gm-ray luminous members of this class and constrain their evolution \citep[e.g.,][]{2022ApJ...931..138F} but also to understand the radiative mechanisms, e.g., leptonic versus hadronic, responsible for the observed high-energy emission.  Indeed, radio galaxies have been proposed as a plausible source of cosmic neutrinos detected with the IceCube observatory \citep[][]{2016JCAP...09..002H} indicating hadronic processes may power their radio jets.

In this Letter, we present the identification and multi-frequency properties of the \gm-ray detected object 4FGL~J1435.5+2021, associated with the radio source TXS~1433+205 \citep[bayesian association probability$\approx$0.99,][]{2020ApJ...892..105A}, as the most distant (as of now) \gm-ray emitting FR~II radio galaxy at $z=0.748$, revealed by the ongoing Very Large Array Sky Survey \citep[VLASS,][]{2020PASP..132c5001L}. This object has been misclassified as a BL Lac in the 4LAC-DR2 catalog though we show that its multi-frequency properties resemble well that of \gm-ray detected FR~II radio galaxies.  We adopt a flat cosmology with $H_0 = 70~{\rm km~s^{-1}~Mpc^{-1}}$ and $\Omega_{\rm M} = 0.3$. At the redshift of $z=0.748$, 1 arcsec$=$7.33 kpc.

\section{Radio Characterization}\label{sec2}
TXS~1433+205 lies in the footprint of several radio surveys, e.g., NRAO VLA Sky Survey \citep[NVSS,][1.4 GHz]{1998AJ....115.1693C},  Faint Images of the Radio Sky at Twenty-Centimeters \citep[FIRST,][1.4 GHz]{1997ApJ...475..479W}, the Galactic and extragalactic all-sky MWA survey \citep[][74$-$231 MHz]{2015PASA...32...25W},  and in the TIFR GMRT Sky Survey \citep[TGSS,][150 MHz]{2017A&A...598A..78I}. However, none of these surveys, except FIRST, could resolve the source morphology.  Even in the FIRST image, at 5 arcsec resolution, TXS~1433+205 is marginally resolved. Between 1.4$-$5 GHz frequencies, the object has a flat radio spectrum \citep[$\alpha=-0.48,~S_\nu\propto\nu^{\alpha}$,][]{2007ApJS..171...61H}. On the other hand, considering the flux densities at 8.4 GHz and 150 MHz, the overall radio spectrum is steep ($\alpha=-0.78$). The 150 MHz luminosity of the source is $\approx5\times10^{27}$ W Hz$^{-1}$, well above the FR I/II threshold.

\begin{figure}
\includegraphics[scale=0.45]{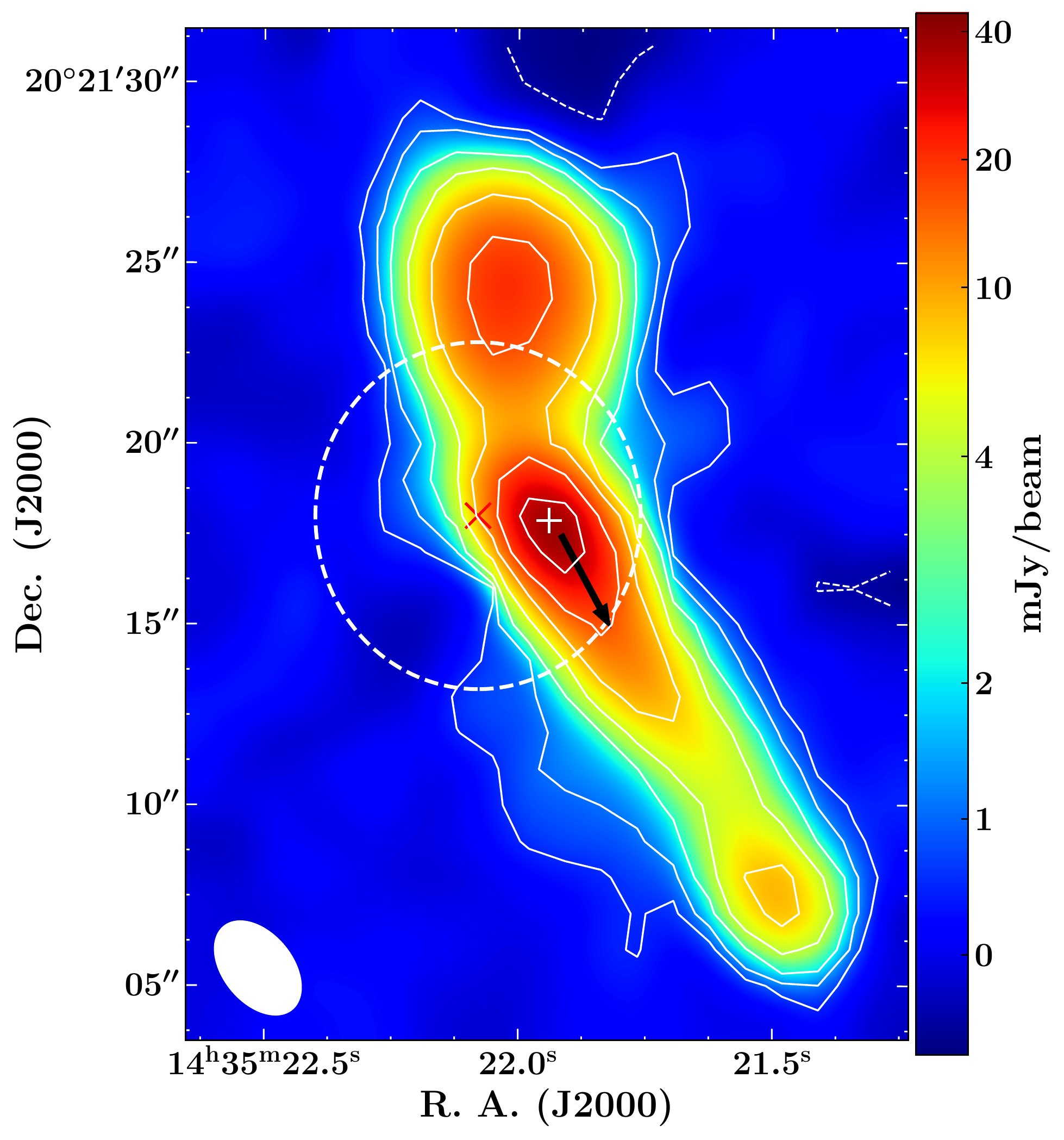}
\caption{VLASS (3 GHz) image of TXS~1433+205. The rms is 0.17 mJy per beam and contours are 3$\times$rms$\times$($-$1, 1, 2, 4, 8, 16, 32, 64). The `X' mark represents the X-ray core position measured by \swift-XRT and the dashed circle refers to the uncertainty in the optimized position at 90\% confidence.  The `+' mark shows the VLBI core position and the black arrow highlights the direction of the boosted jet. North is up and east to the left.} \label{fig:1}
\end{figure}

 VLASS is an ongoing radio survey at 2$-$4 GHz with high angular resolution ($\sim$2.5 arcsec) and sensitivity (0.12 mJy per beam for the first VLASS epoch).  We browsed the VLASS quick look images\footnote{https://science.nrao.edu/vlass/data-access/vlass-epoch-1-quick-look-users-guide} of TXS~1433+205 and found a resolved radio morphology. As can be seen in Figure~\ref{fig:1}, the object is resolved into three components besides the extended radio jet: a central core and two lobes with hotspots. These radio observations, along with the optical (Section~\ref{sec5}) and multi-wavelength (Section~\ref{sec3}) observations, unambiguously confirm that TXS~1433+205 is a \gm-ray emitting FR~II radio galaxy.  The radio core of the object is positionally consistent with a moderately bright X-ray source detected with the \swift~X-ray Telescope (XRT, observed $F_{\rm 0.3-10~keV}\sim3\times10^{-13}$\ergflux). The north radio lobe ends at $\sim$10 arcsec (projected length $\sim$73 kpc) far from the core, whereas, the south-west lobe at a distance of $\sim$14 arcsec (projected length $\sim$104 kpc). 

The flux density of the core is comparable to that of the northern lobe,  57.7$\pm$0.4 mJy, and 67.2$\pm$0.6 mJy, respectively, whereas the south-western lobe is considerably fainter with a flux density of 12.8$\pm$0.4 mJy.  The peak flux densities of the northern and south-western hotspots are 21.3$\pm$0.2 mJy/beam and 10.3$\pm$0.2 mJy/beam, respectively.
The Laing-Garrington effect clearly demonstrated that the apparent asymmetry of jets in radio-loud AGN is due to relativistic beaming with the prominent jet approaching us \citep{1988Natur.331..147G,1988Natur.331..149L}. Hence the hotspot and lobe facing the jet is also on the approaching side and usually tends to be brighter than the one on the opposite side, possibly due to mild relativistic beaming of the hotspots \citep[e.g.][]{1991MNRAS.250..171G}. In the case of  TXS~1433+205, a nuclear Very long baseline interferometry (VLBI)-scale jet is also seen to point in a similar direction as the large-scale jet. The VLBI image taken in the X-band \citep[]{2021AJ....161...14P} is shown in the appendix (Figure~\ref{fig:vlbi}).
The considerably fainter hotspot and lobe on its south-western side, which face the jet and hence are on the approaching side, suggest significant dissipation of energy in the jet. Several such examples are known in the literature \citep[for a review see e.g.][]{2022arXiv220605803S}.
Furthermore, we compute the core dominance ($C_{\rm D}$) following \citet[][]{2010ApJ...720..912A}. It is determined as $C_{\rm D}=\log(F_{\rm core}/(F_{\rm tot}-F_{\rm core}))$, where $F_{\rm tot/core}$ is the total/core flux density in the source rest-frame at 3 GHz. This parameter allows us to constrain the viewing angle of the jet since $F_{\rm core}\propto\delta^{3+\alpha}$ \citep[$\delta$ is the Doppler factor, cf.][]{1995ApJ...446L..63D} whereas the lobe emission is expected to be largely isotropic and hence unbeamed. For TXS~1433+205, we get $C_{\rm D}=-0.11$ which is similar to other \gm-ray detected FR~I and II radio galaxies \citep[][]{2010ApJ...720..912A}. This indicates a relatively small viewing angle of the jet. Indeed, large core dominance hints that AGNs viewed closer to the jet axis have a higher probability to be detected with the \fermi-LAT.

\begin{figure}
\vbox{
\includegraphics[width=\linewidth]{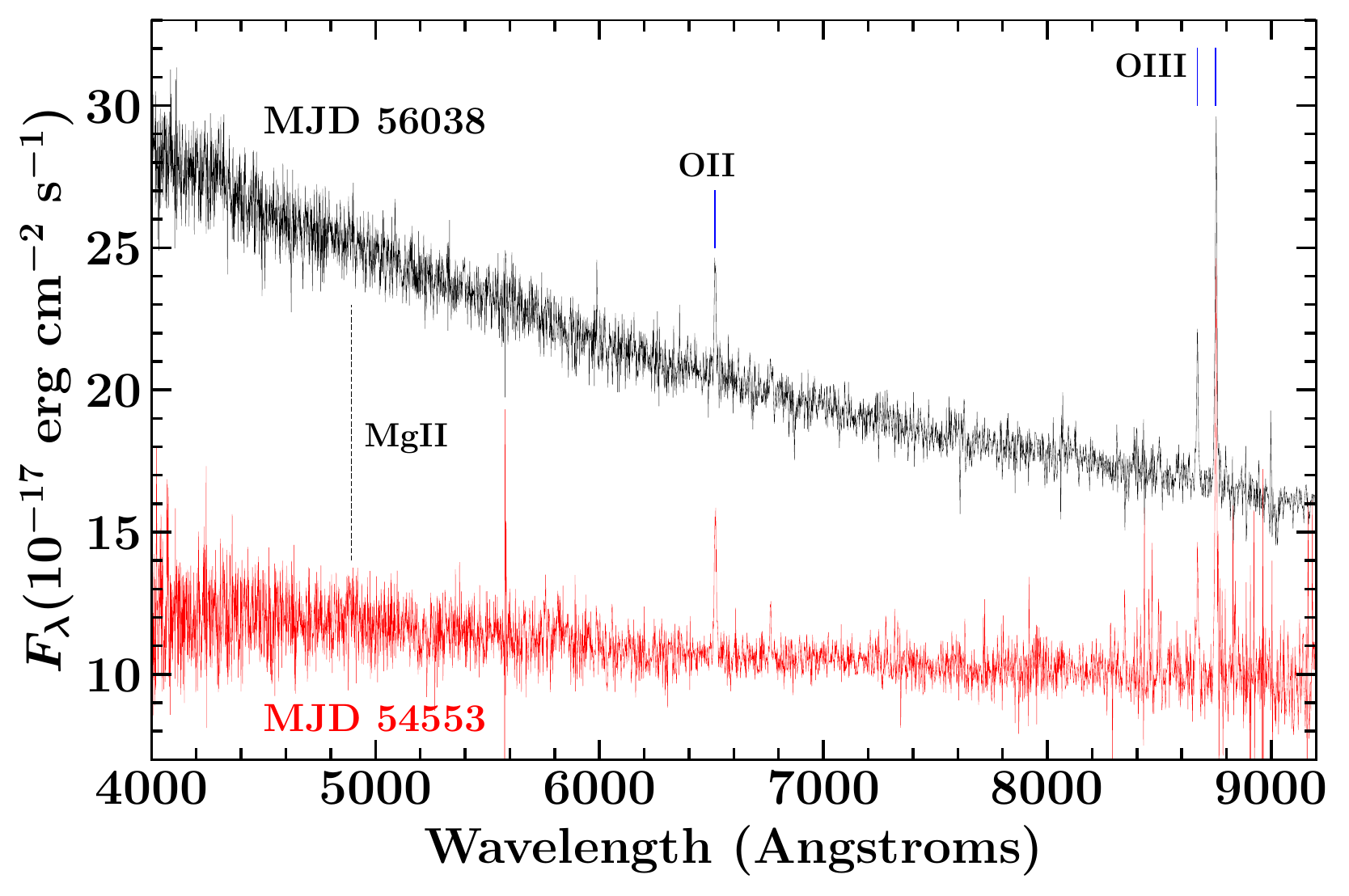}
\includegraphics[width=\linewidth]{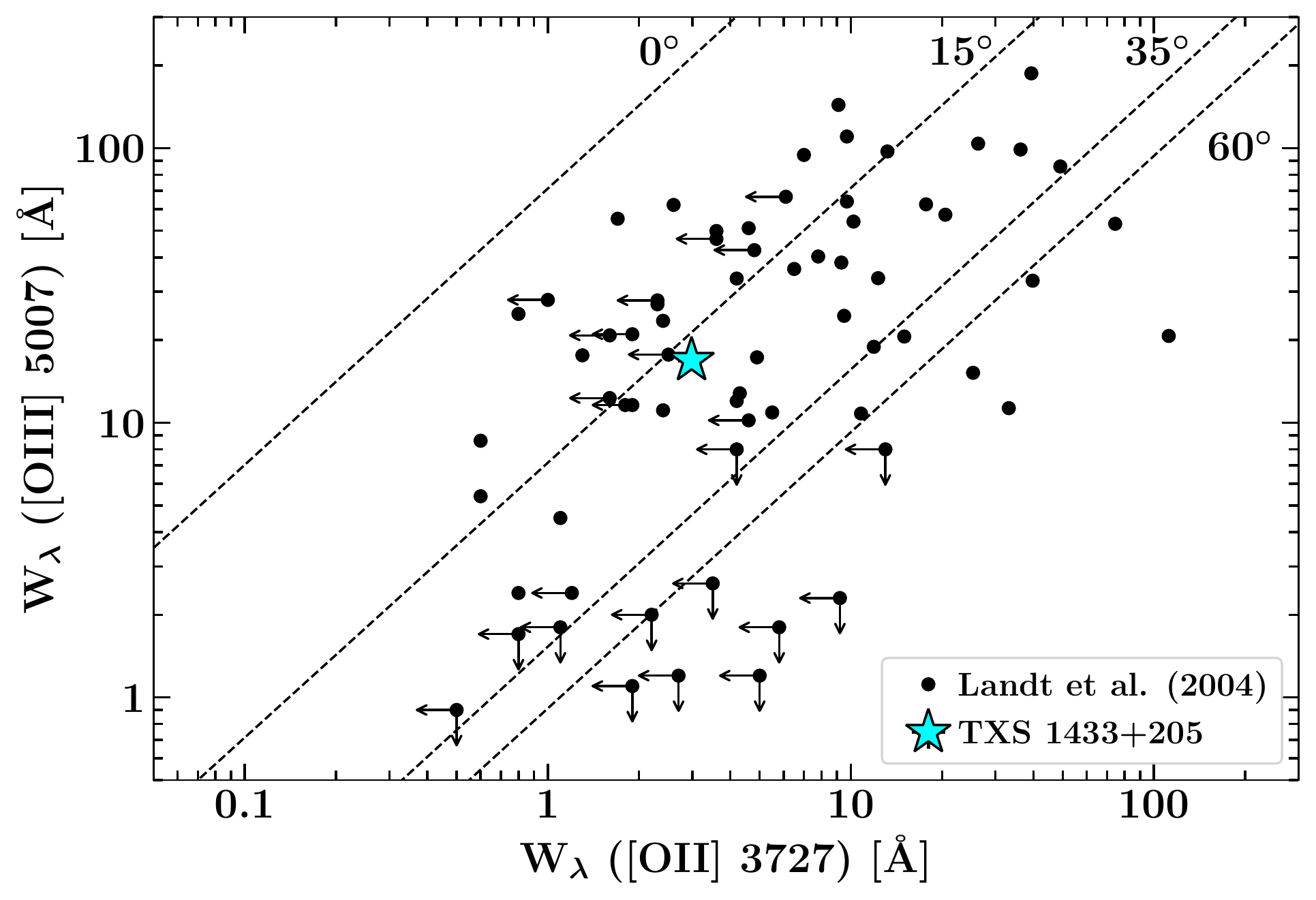}
\includegraphics[width=\linewidth]{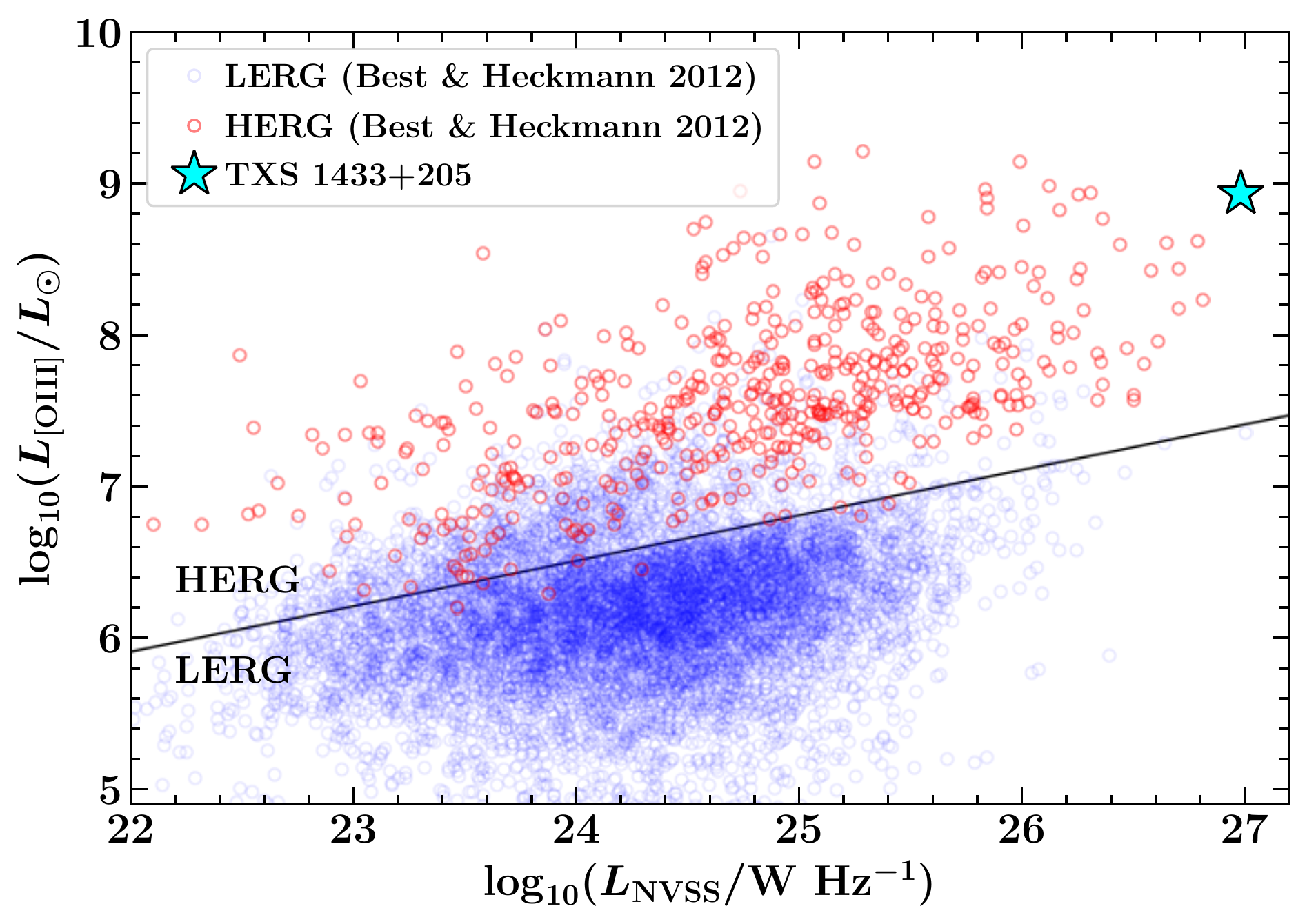}
}
\caption{Top: The SDSS optical spectrum of TXS~1433+205. Prominent emission/absorption features are marked. Middle: This plot shows the position of TXS~1433+205 in the [O~{\sc ii}]$\lambda$3727 and [O~{\sc iii}]$\lambda$5007 EW plane. The shown lines, highlighting the jet viewing angles, and other jetted sources (black circles) are adopted from \citet[][]{2004MNRAS.351...83L}. Bottom: The location of TXS 1433+205 in the [O~{\sc iii}] emission line luminosity versus radio luminosity diagram. The solid line indicates an approximate divide between HERGs and LERGs proposed by \citet[][]{2012MNRAS.421.1569B}. } \label{fig:2}
\end{figure}

Recently, \citet[][]{2022MNRAS.514.2122P} have reported the identification of seven blazars with extended jet morphologies similar to radio galaxies using MHz frequency observations taken with the Low-Frequency Array (LOFAR). The change in the direction of the jet out of the sky plane possibly due to some kind of precession mechanism may lead to the observation of extended radio emission revealing the older AGN activity as proposed by \citet[][]{2022MNRAS.514.2122P}.  The LOFAR observations taken at 144 MHz probably trace the old/previous AGN phase since the high-energy electrons emitting GHz frequency radiation cool faster than those radiating MHz frequency emission. Since TXS~1433+205 exhibits bi-polar extended emission at GHz frequencies as observed by VLASS and a normal low-frequency spectral index of $-$0.78, it probably represents the current jet activity. The overall structure appears misaligned which is supported by the fact that the Northern lobe lies on a different axis with respect to the one connecting the core and Southern lobe. This could also be due to interaction with the inter-galactic medium, which could contribute to its shorter distance from the core, and the stronger flux density\footnote{Part of the misalignment could be due to projection effects as the source is viewed at a relatively small angle to the line of sight, although not head-on.}. It is not covered in the second data release of the LOFAR survey \citep[][]{2022A&A...659A...1S} and is not resolved in the 888 MHz observations taken with the Rapid ASKAP Continuum Survey \citep[RACS,][]{2020PASA...37...48M}. High-resolution, low-frequency MHz observations of TXS~1433+205 will be required to reveal the episodic AGN activity, if any.

\section{Optical Spectroscopic Classification}\label{sec5}
The optical spectrum of TXS~1433+205 was first published by \citet[][]{1996MNRAS.282.1274H} and found to be featureless.  This may have led to its classification as a BL Lac object \citep[][]{2010A&A...518A..10V}. However, better quality spectra taken with the Sloan Digital Sky Survey (SDSS, Figure~\ref{fig:2}, top panel) revealed the presence of strong [O~{\sc ii}]$\lambda$3727 and [O~{\sc iii}]$\lambda$5007 emission lines securing the source redshift as $z=0.748$. This rules out the BL Lac classification of the source. Recently, \citet[][]{2022Univ....8..587F} have also updated the classification of TXS~1433+205 as a misaligned AGN based on the observation of [O~{\sc ii}]$\lambda$3727 and [O~{\sc iii}]$\lambda$5007 emission lines and steep radio spectrum as also found by us. Interestingly, SDSS spectra taken at two different epochs show `bluer-when-brighter' flux variations which can be explained due to the non-negligible contamination by the jet. The low-activity state spectrum, on the other hand, hints for the presence of broad {Mg~\sc ii} and H$_{\beta}$ emission lines although the signal-to-noise ratio of the data is poor\footnote{\url{https://tinyurl.com/9adwdtjf}}.

\citet[][]{2004MNRAS.351...83L} proposed that beamed AGN and radio galaxies can be separated on the [O~{\sc ii}]$\lambda$3727$-$[O~{\sc iii}]$\lambda$5007 equivalent width (EW) plane, thus constraining the viewing angle of the jet. This is because the line radiation is expected to be isotropic and the AGN continuum and jet contamination vary with the viewing angle, hence the line EWs as well \citep[see also,][]{2011MNRAS.411.2223R}. Since both [O~{\sc ii}]$\lambda$3727 and [O~{\sc iii}]$\lambda$5007 lines are strong, we directly use the EW measured by the SDSS pipeline \citep[][]{2020ApJS..249....3A} and found it to be $2.98\pm0.30$ \AA~and $16.87\pm0.61$ \AA, respectively. In Figure~\ref{fig:2} (middle panel), we show the position of TXS~1433+205 in the line EW diagram \citep[][]{2004MNRAS.351...83L} and estimate the viewing angle to be $\sim$15$^{\circ}$. The presence of such a moderately misaligned jet is supported by the \gm-ray detection and flux variability seen in the SDSS spectrum.

\begin{figure*}
\hspace{0.cm}\hbox{
\includegraphics[scale=0.4]{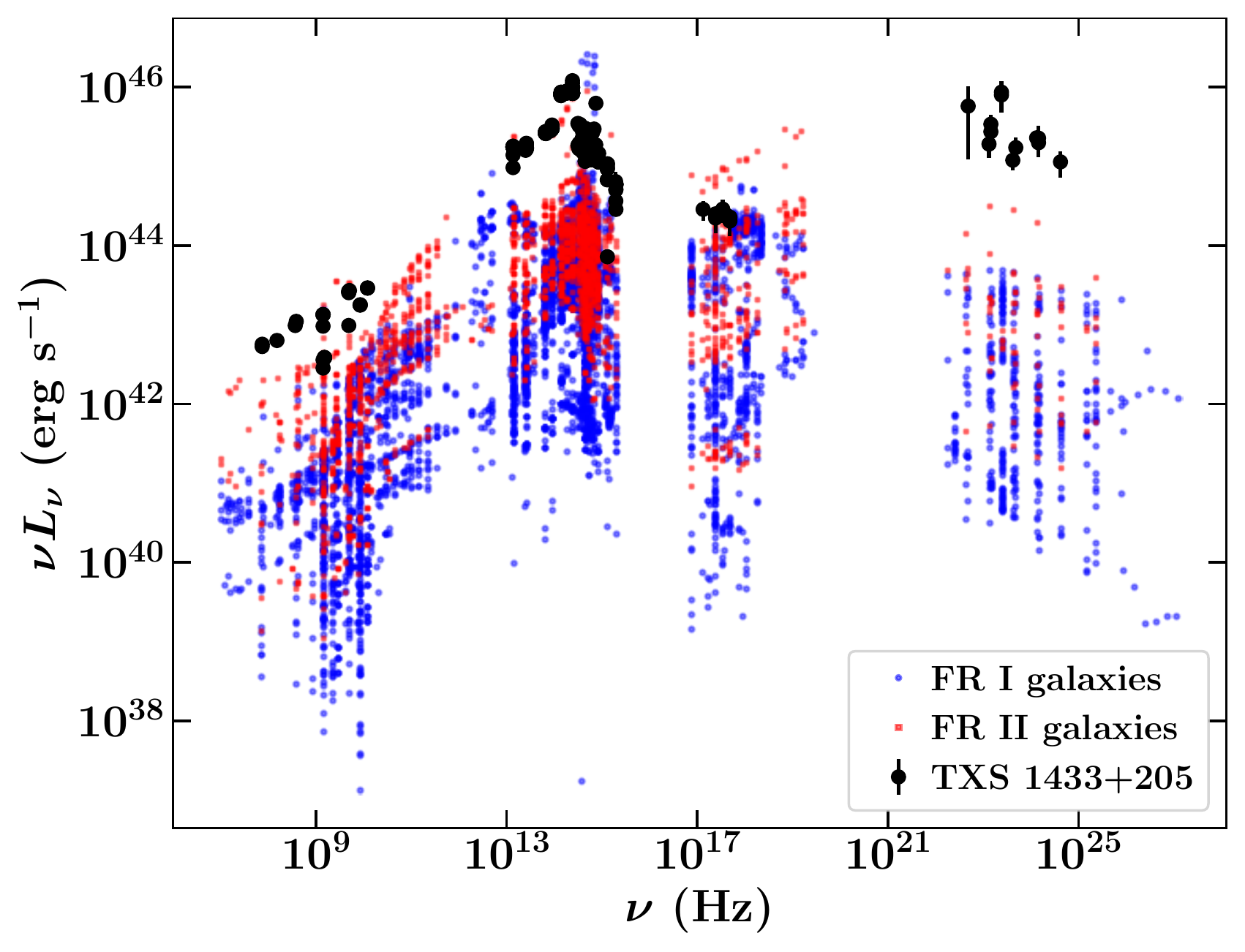}
\hspace{1.0cm}
\includegraphics[scale=0.4]{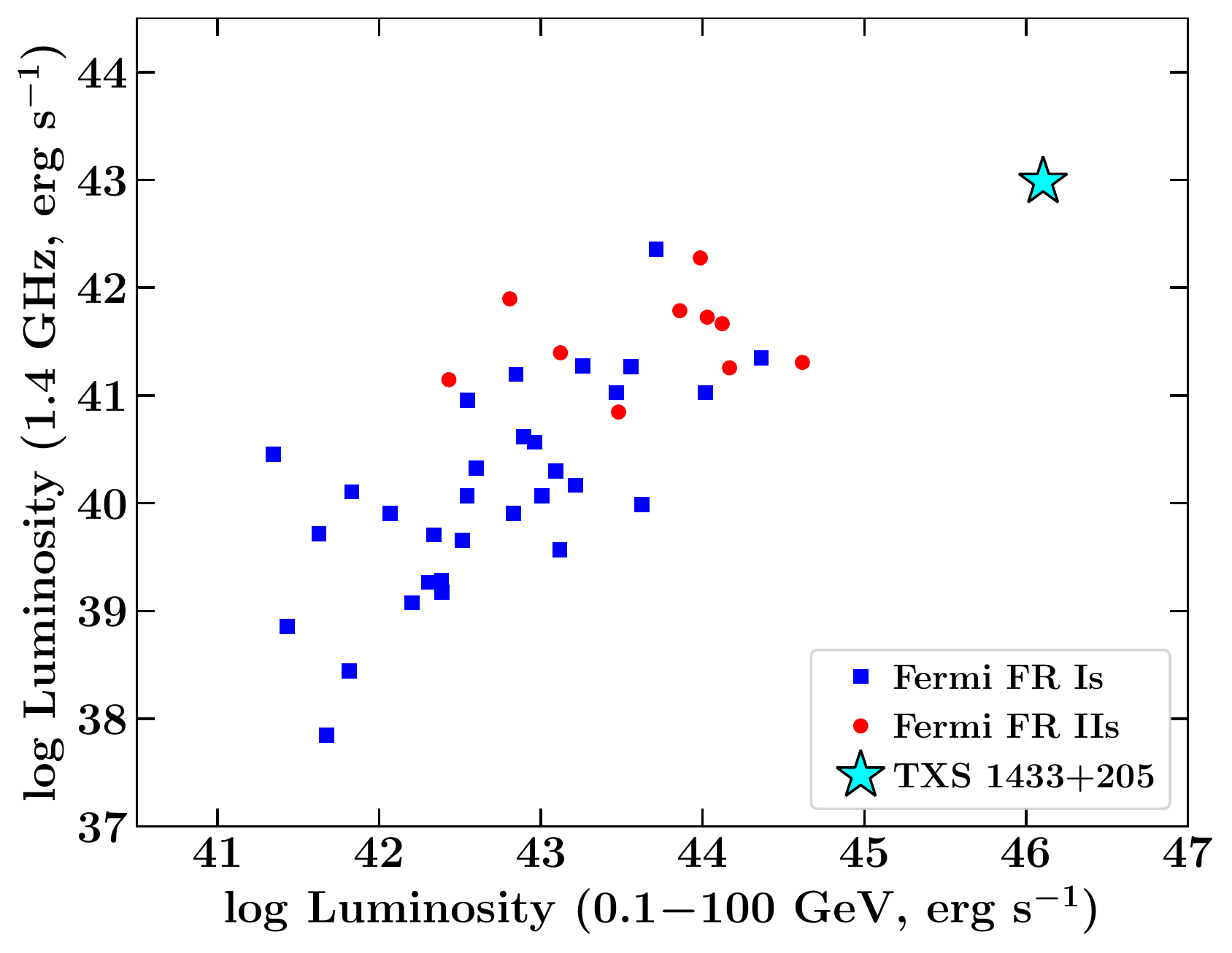}
}
\hspace{1.3cm}\hbox{
\includegraphics[scale=0.4]{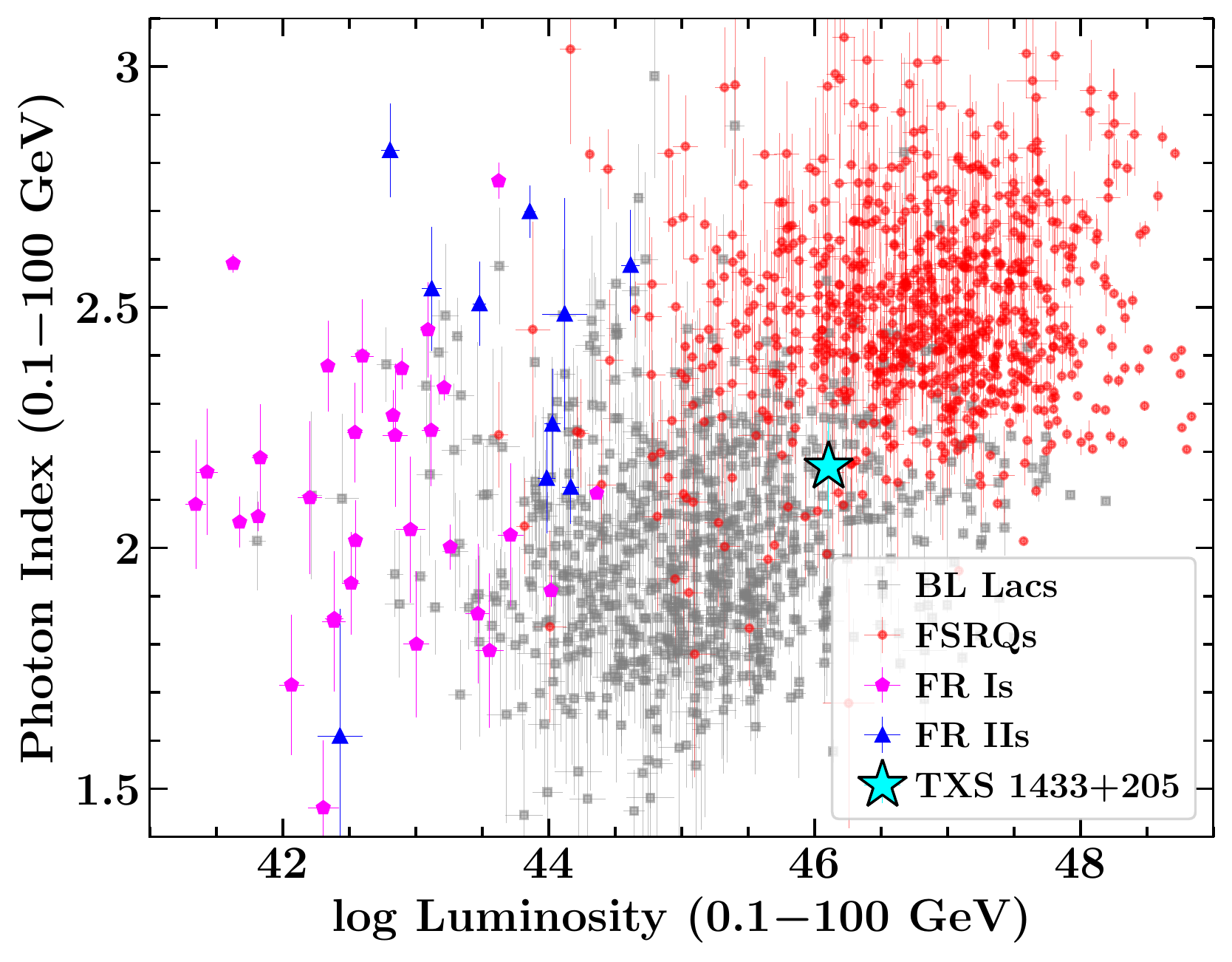}
\hspace{1.0cm}
\includegraphics[scale=0.4]{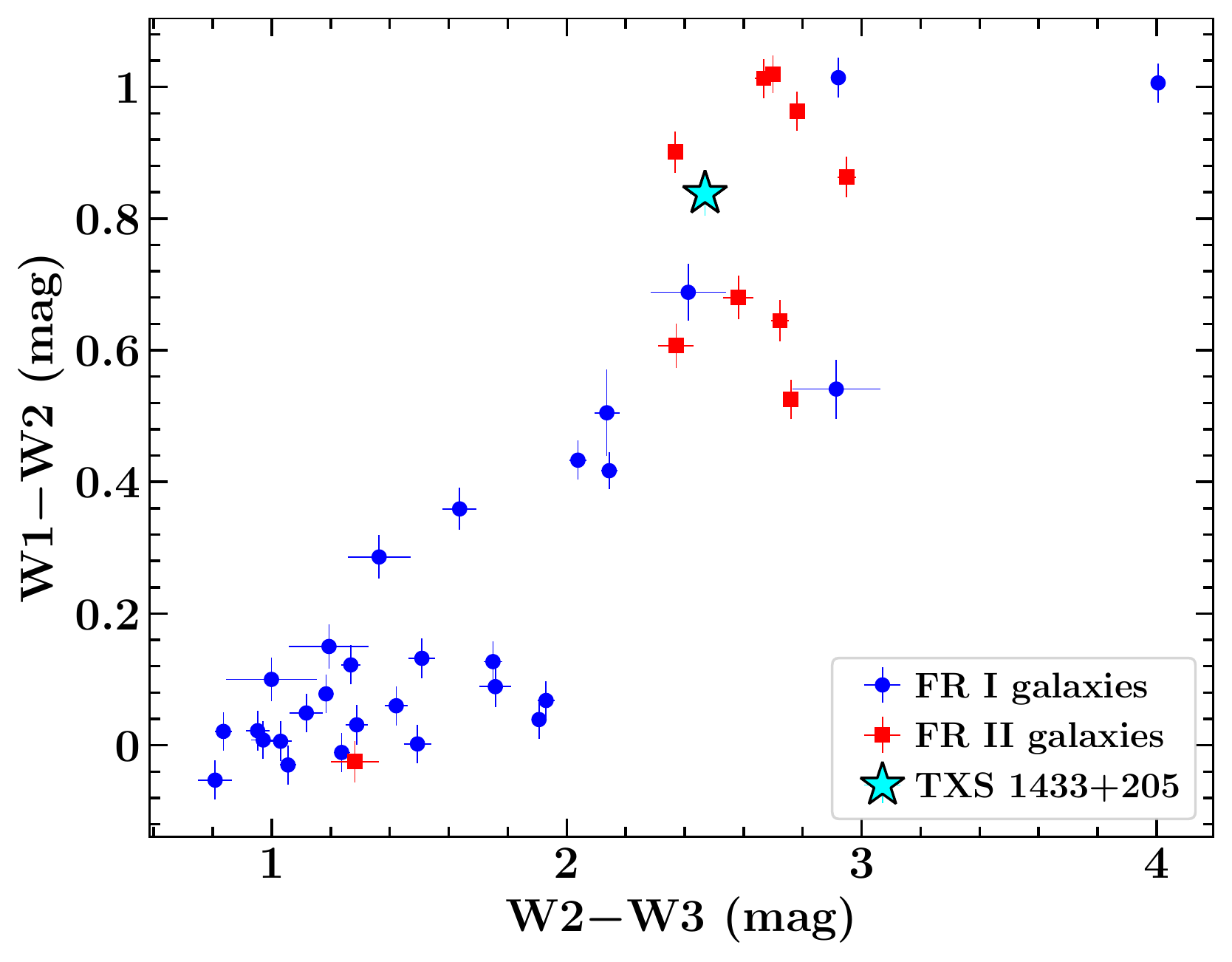}
}
\caption{Top left: The multi-wavelength SED of TXS~1433+205 plotted along with other \gm-ray detected FR~I and II radio galaxies.  Top right: This plot shows the radio luminosity as a function of the \gm-ray luminosity for \fermi-LAT detected radio galaxies. Bottom left: The variation of the \gm-ray spectral shape as a function of the \gm-ray luminosity. Bottom right: The WISE colour diagram of \gm-ray emitting radio galaxies.} \label{fig:3}
\end{figure*}

To determine whether TXS~1433+205 belongs to the class of the high-excitation or low-excitation radio galaxies (HERG and LERG, respectively), we highlight its position in the radio-[O~{\sc iii}]$\lambda$5007 luminosity diagram (Figure~\ref{fig:2}, bottom panel).  For comparison, we also show the galaxies studied by \citet[][]{2012MNRAS.421.1569B} and their proposed HERG/LERG divide. In this diagram, TXS~1433+205 lies in a region mainly populated by HERGs and well above the HERG/LERG division line which indicates a radiatively efficiently accretion \citep[accretion rate $>1\%$ of the Eddington rate,][]{2012MNRAS.421.1569B} powering the system. A large EW ($>$3 \AA) of the [O~{\sc iii}]$\lambda$5007 line also supports the HERG classification of the source \citep[][]{1997MNRAS.286..241J,2009A&A...495.1033B}. 

\section{Multi-wavelength Properties}\label{sec3}
The top left panel of Figure~{\ref{fig:3}} shows the broadband spectral energy distribution (SED) of TXS~1433+205. For comparison, we also plot the SEDs of 42 \fermi-LAT detected FR~I and II sources \citep[][]{2020ApJ...892..105A,2022ApJ...931..138F} using the data taken from the Space Science Data Center SED builder tool\footnote{https://tools.ssdc.asi.it/SED}. A quick look at this plot suggests TXS~1433+205 to be among the most luminous \gm-ray detected radio galaxies. In particular, this object is the most luminous in the radio and \gm-ray bands as can be seen in Figure~{\ref{fig:3}} (top right panel).  Compared to blazars, the 0.1$-$100 GeV isotropic \gm-ray luminosity of TXS~1433+205 is similar to the low-luminosity flat spectrum radio quasars but is on the higher side of that for BL Lac objects (Figure~{\ref{fig:3}}, bottom left panel). The \gm-ray spectrum of the source (photon index $\Gamma_\gamma=2.17\pm0.09$) appears harder in comparison to other FR~II radio galaxies ($\langle\Gamma_\gamma\rangle=2.38\pm0.05$). This observation can explain the detection of such a distant radio galaxy with the \fermi-LAT. 

The 0.3$-$10 keV X-ray spectrum of TXS~1433+205 is steep ($\Gamma_{\rm X}=2.18\pm0.68$) in contrast to other FR~II radio galaxies (Figure~{\ref{fig:3}}, top left panel) that typically exhibit flat X-ray spectra 
\citep[$\Gamma_{\rm X}<2$, cf.][]{2022ApJ...931..138F}.  Such a steep falling X-ray spectrum is observed from the high synchrotron peaked BL Lac objects and is explained as the tail of the synchrotron emission \citep[e.g.,][]{2011ApJ...736..131A}. If so, the corresponding inverse Compton peak is expected to be located at very high energies ($>$10 GeV) leading to the observation of a rising \gm-ray spectrum ($\Gamma_\gamma<2$). However, a soft \gm-ray spectrum is detected from TXS~1433+205, thus challenging the conventional one-zone leptonic emission scenario. One possibility could be that there could be a non-negligible contribution from the X-ray emitting corona in the 0.3$-$10 keV band \citep[e.g.,][]{2007ApJ...659..235G}.  Alternatively, more complex radiative processes, e.g., multi-zone structured jet \citep[][]{2002ApJ...578..763S}, could be responsible for the observed peculiar X-ray and \gm-ray spectra.

The infrared (IR)-to-ultraviolet SED of TXS~1433+205 shows a peak, likely due to the host galaxy, similar to that observed from other \gm-ray detected radio galaxies. In the {\it Wide-field Infrared Survey Explorer} (WISE) colour-colour diagram, the source lies in a region mainly populated by \gm-ray detected FR~II radio galaxies (Figure~{\ref{fig:3}}, bottom right panel). Furthermore, to determine the contribution of the star-formation activities, we calculated the two-point spectral index $\alpha^{22}_{\rm 1.4}$ using the rest-frame flux densities at 22 $\mu$m and 1.4 GHz, as follows:

\begin{equation}
\alpha^{22}_{\rm 1.4}=\frac{\log(F_{\rm 1.4~GHz}/F_{\rm 22~\mu m})}{\log(\nu_{\rm 1.4~GHz}/\nu_{\rm 22~\mu m})}
\end{equation}

The typical value of $\alpha^{22}_{\rm 1.4}$ for jetted AGNs is $>0.2$ \citep[cf.][]{2015MNRAS.451.1795C}. On the other hand, for sources with $\alpha^{22}_{\rm 1.4}<-0.25$, star-formation activities are expected to dominate the observed radio emission. Using NVSS and WISE data, we found $\alpha^{22}_{\rm 1.4}=0.47$ for TXS~1433+205. This indicates the dominance of the jet emission in the observed radio and IR emission.

With both VLASS and \fermi-LAT operations active, along with other ongoing surveys, e.g., LOFAR and RACS, the identification of TXS~1433+205 as a \gm-ray emitting radio galaxy can be considered only as the tip of the iceberg \citep[see also][]{2022MNRAS.513..886B}.  Among the known \gm-ray detected objects, the advent of the Square Kilometer Array will enable us to make deep images of these objects, and along with multi-wavelength data, may lead to the re-classification of several blazars or blazar candidates as radio galaxies. At the same time, the continuous monitoring of the \gm-ray sky with the \fermi-LAT and optical spectroscopic followups will be crucial to detect faint \gm-ray emitting AGN and constraining the high-energy radiation mechanism of these enigmatic cosmic radio sources.

\section*{Acknowledgements}
We are grateful to the referee, Gabriele Bruni, for constructive criticism, which has helped improve the paper. VSP acknowledges a useful discussion with R. Srianand (IUCAA) and N. Gupta (IUCAA). We acknowledge the use of the SDSS and WISE data. Part of this work is based on archival data, software or online services provided by the Space Science Data Center - ASI. The National Radio Astronomy Observatory is a facility of the National Science Foundation operated under cooperative agreement by Associated Universities, Inc. CIRADA is funded by a grant from the Canada Foundation for Innovation 2017 Innovation Fund (Project 35999), as well as by the Provinces of Ontario, British Columbia, Alberta, Manitoba and Quebec.

\section*{Data Availability}
All of the multi-wavelength data used in this article are publicly available at their respective data archives, e.g., VLASS cutout server (http://cutouts.cirada.ca).


\bibliographystyle{mnras}
\bibliography{Master} 

\appendix

\section{VLBI Image of TXS 1433+205}

Here we show the VLBI image of TXS 1433+205 taken from the list of VLBI Calibrators\footnote{\url{http://astrogeo.org/calib/search.html}} \citep[cf.][]{2021AJ....161...14P}.

\begin{figure}
\includegraphics[scale=0.5, clip, trim={0 0 1cm 9.5cm}]{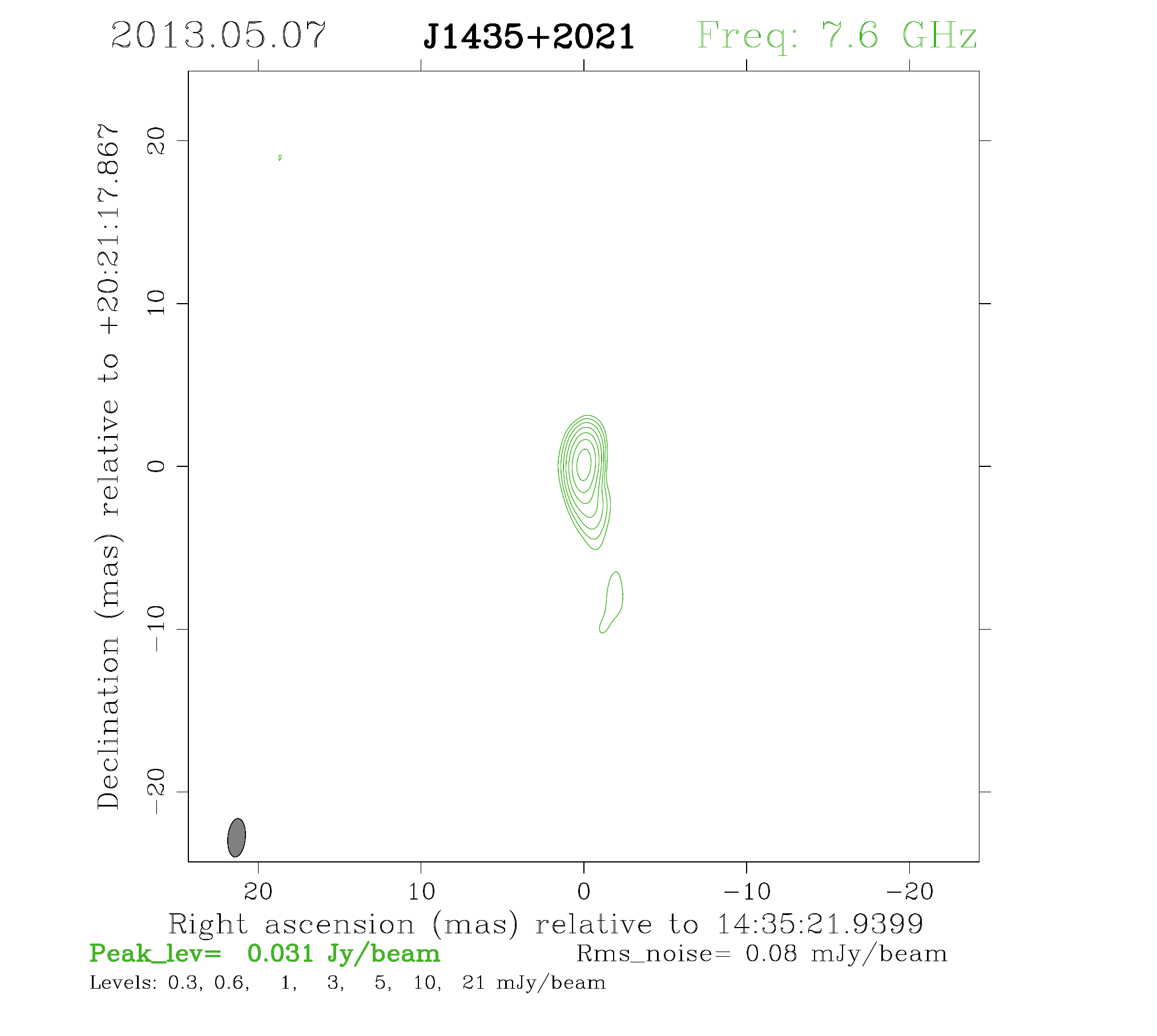}
\caption{VLBI image of TXS 1433+205. Other information are provided in the figure.} \label{fig:vlbi}
\end{figure}


\bsp	
\label{lastpage}
\end{document}